\pdfoutput=1

\documentclass[11pt]{article}
\usepackage{algorithm}
\usepackage{algpseudocode}
\usepackage{colortbl}  
\usepackage{xcolor}    
\usepackage{pifont}    
\usepackage{makecell}  
\definecolor{GrayHeader}{gray}{0.9}
\definecolor{CheckGreen}{rgb}{0.1, 0.6, 0.1}
\definecolor{CrossRed}{rgb}{0.8, 0.0, 0.0}
\newcommand{\cmark}{\textcolor{CheckGreen}{\ding{51}}}
\newcommand{\xmark}{\textcolor{CrossRed}{\ding{55}}}
\usepackage[final]{acl}
\usepackage{kotex}
\usepackage{times}
\usepackage{latexsym}
\usepackage[T1]{fontenc}
\usepackage[utf8]{inputenc}
\usepackage{microtype}
\usepackage{inconsolata}
\usepackage{graphicx}
\usepackage{enumitem}
\usepackage{booktabs}  
\usepackage{multirow}
\usepackage{adjustbox}  
\usepackage{listings}   
\definecolor{LightBlue}{RGB}{230, 242, 255}  
\definecolor{CodeBg}{RGB}{245, 245, 245}     

\definecolor{ShiftColor}{RGB}{66, 133, 244}    
\definecolor{OracleColor}{RGB}{52, 168, 83}    
\definecolor{LiveColor}{RGB}{251, 188, 5}      
\definecolor{InterColor}{RGB}{234, 67, 53}     
\definecolor{DefensiveColor}{RGB}{156, 39, 176}
\definecolor{SafetyColor}{RGB}{96, 125, 139}   

\usepackage[most]{tcolorbox}

\newtcolorbox{promptbox}[2][]{%
    enhanced,
    colback=#2!5!white,
    colframe=#2!75!black,
    fonttitle=\bfseries\sffamily,
    attach boxed title to top left={yshift=-2mm, xshift=3mm},
    boxed title style={colback=#2!75!black, colframe=#2!75!black},
    top=4mm,
    boxrule=0.5pt,
    #1
}

\newtcolorbox{codebox}[1][]{%
    enhanced,
    colback=CodeBg,
    colframe=gray!50!black,
    fonttitle=\bfseries\sffamily,
    attach boxed title to top left={yshift=-2mm, xshift=3mm},
    boxed title style={colback=gray!60, colframe=gray!50!black},
    top=4mm,
    boxrule=0.4pt,
    left=2mm, right=2mm,
    #1
}

\newtcolorbox{safetybox}[1][]{%
    enhanced,
    colback=SafetyColor!8!white,
    colframe=SafetyColor!80!black,
    fonttitle=\bfseries\sffamily,
    attach boxed title to top left={yshift=-2mm, xshift=3mm},
    boxed title style={colback=SafetyColor!80!black, colframe=SafetyColor!80!black},
    top=4mm,
    boxrule=0.5pt,
    #1
}

\newtcolorbox{examplebox}[1][]{%
    enhanced,
    breakable,
    colback=gray!3!white,
    colframe=black!70,
    fonttitle=\bfseries\sffamily\large,
    attach boxed title to top center={yshift=-3mm},
    boxed title style={colback=black!70, colframe=black!70},
    top=6mm,
    boxrule=0.6pt,
    left=3mm, right=3mm,
    before upper={\parindent0pt\small},
    #1
}

\newtcblisting{codelistbox}[2][]{%
    enhanced,
    colback=#2!5!white,
    colframe=#2!80!black,
    fonttitle=\bfseries\sffamily,
    attach boxed title to top left={yshift=-2mm, xshift=3mm},
    boxed title style={colback=#2!80!black, colframe=#2!80!black},
    top=2mm, bottom=2mm,
    boxrule=0.4pt,
    left=2mm, right=2mm,
    listing only,
    listing options={
        language=Python,
        basicstyle=\ttfamily\footnotesize,
        breaklines=true,
        columns=flexible,
        keepspaces=true,
        showstringspaces=false,
        commentstyle=\color{gray},
        keywordstyle=\color{blue!70!black},
        stringstyle=\color{orange!80!black},
        morekeywords={self, True, False, None},
    },
    #1
}

\title{SolidCoder: Bridging the Mental-Reality Gap in LLM Code Generation through Concrete Execution}

\author{
Woojin Lee \quad
Jin-Xia Huang \\
\\Electronics and Telecommunications Research Institute, Republic of Korea  \\
\texttt{\{writerwoody, hgh\}@etri.re.kr} \\
} 

\begin{document}
\maketitle
\begin{abstract}
State-of-the-art code generation frameworks rely on \textit{mental simulation}, where LLMs internally trace execution to verify correctness. We expose a fundamental limitation: the \textbf{Mental-Reality Gap}---where models hallucinate execution traces and confidently validate buggy code. This gap manifests along two orthogonal dimensions: the \textbf{Specification Gap} (overlooking edge cases during planning) and the \textbf{Verification Gap} (hallucinating correct behavior for flawed code). We propose \textbf{SolidCoder} with a simple principle: \textit{don't imagine---execute}. The \textbf{S.O.L.I.D.} architecture addresses both dimensions by forcing edge-case awareness before algorithm design and replacing imagined traces with sandboxed execution using property-based oracles. With GPT-4o, SolidCoder achieves state-of-the-art pass@1 performance: \textbf{95.7\%} on HumanEval (+0.6\%p), \textbf{77.0\%} on CodeContests (+4.3\%p), and \textbf{26.7\%} on APPS (+3.4\%p). Ablation reveals that edge-case awareness provides the largest individual gain, while execution grounding catches categorically different errors that specification improvements cannot address. These gains generalize to RL post-trained models, validating that bridging both gap dimensions is essential for robust code synthesis. We release our code and framework to facilitate future research.\footnote{\url{https://github.com/10kH/SolidCoder}}
\end{abstract}

\section{Introduction}
\label{sec:intro}

\begin{figure*}[t]
    \centering
    \includegraphics[width=\textwidth]{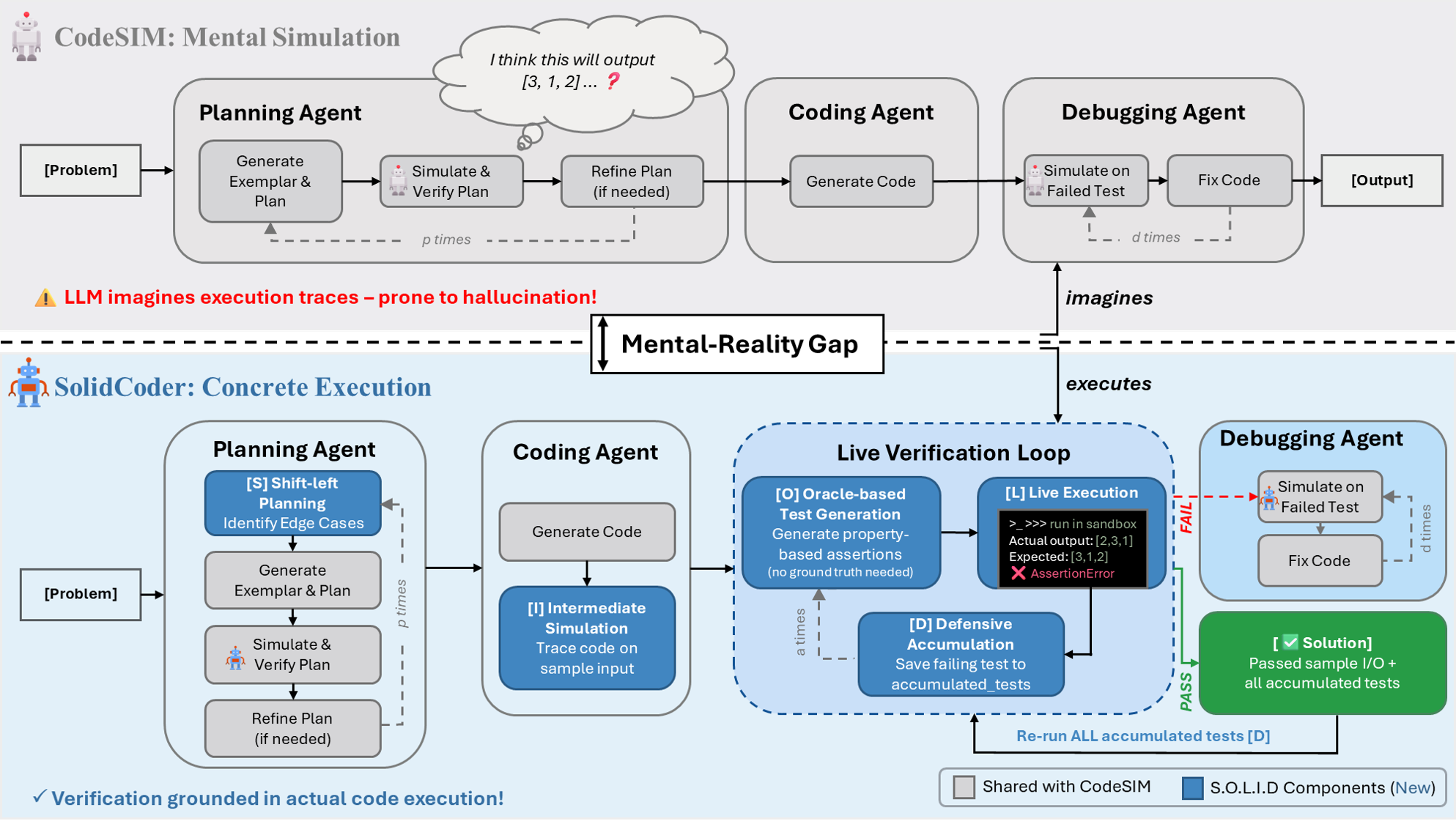}
    \caption{\textbf{Comparative Architecture Overview: CodeSIM vs. SolidCoder.} The top pipeline illustrates the CodeSIM baseline, which relies on mental simulation ("imagines") for verification during planning and debugging, resulting in a "Mental-Reality Gap" prone to LLM hallucination. The bottom pipeline demonstrates SolidCoder, which bridges this gap by grounding verification in concrete execution ("executes"). SolidCoder incorporates S.O.L.I.D. enhancements highlighted in \textcolor{blue}{blue}: (1) The \textbf{Planning Agent} integrates \textbf{S}hift-left Planning to identify edge cases before plan generation; (2) The \textbf{Coding Agent} applies \textbf{I}ntermediate Simulation to verify code traces; (3) The central \textbf{Live Verification Loop} replaces mental checks with sandbox \textbf{L}ive Execution utilizing \textbf{O}racle-based assertions. Failing tests are stored via \textbf{D}efensive Accumulation, and fixed code must pass *all* accumulated tests ([D]) to prevent regression before being accepted as a solution.}
    \label{fig:1}
\end{figure*}

Recent large language models have demonstrated remarkable programming capabilities \citep{ge2025survey}. Foundation model families such as GPT \citep{achiam2023gpt}, Gemini \citep{team2023gemini}, Claude \citep{anthropic2024claude3}, DeepSeek \citep{liu2025deepseek}, Kimi \citep{team2025kimi}, and Qwen \citep{yang2025qwen3} have shown impressive proficiency in generating code from natural language descriptions. However, synthesizing correct programs remains inherently challenging, requiring deep understanding of natural language processing, algorithmic design, data structure selection, and problem-solving strategies. These challenges are compounded for competitive programming and advanced software engineering tasks, where adherence to strict constraints and passing comprehensive test suites are paramount \citep{chen2021evaluating, hendrycks2021measuring, li2022competition, jimenez2023swe, jain2024livecodebench}.

To overcome these challenges, a rich line of research has emerged. Early methods employed direct prompting \citep{chen2021evaluating}, Chain-of-Thought reasoning \citep{wei2022chain}, retrieval-augmented generation \citep{parvez2021retrieval}, and various in-context exemplar-based strategies \citep{zhang2022automatic, shum2023automatic}. Recent paradigms have shifted toward plan-based decomposition \citep{jiang2024self}, tree-searching approaches \citep{zhou2023language}, analogical reasoning \citep{yasunaga2023large}, and diverse agent-based architectures \citep{shinn2023reflexion, qian2024chatdev}. Among these, MapCoder \citep{islam2024mapcoder} marked a significant advance by introducing a multi-agent pipeline that separates concerns: one agent retrieves analogous solved problems, another constructs execution plans, a third generates code, and a fourth handles debugging through iterative refinement. Building on this foundation, CodeSIM \citep{islam2025codesim} achieved state-of-the-art results by leveraging \textit{Mental Simulation} at two critical phases: (1) in the planning agent, the LLM simulates execution to verify the plan before code generation, and (2) in the debugging agent, when code fails tests, simulation is employed again to trace through the logic and identify bugs.

However, this approach has a fundamental flaw: \textbf{LLMs hallucinate}. In complex algorithmic scenarios, models frequently ``imagine'' execution traces that do not reflect actual program behavior---effectively engaging in wishful thinking where they convince themselves that flawed code works correctly \citep{huang2025survey}. This is akin to playing blindfold chess and declaring victory without ever moving the pieces on the board. We term this phenomenon the \textbf{``Mental-Reality Gap''}---the disconnect between how an agent \textit{believes} a program behaves and how it \textit{actually} behaves at runtime. This gap causes agents to confidently validate buggy code, submitting incorrect solutions as final answers. Worse, such a self-affirming verification loop can reinforce errors rather than correcting them.

The most direct way to bridge the Mental-Reality Gap is to \textit{actually execute} the code rather than imagining its behavior. Two barriers have historically prevented this approach. The first is \textbf{technical feasibility}: when LLMs were pure text generators, interaction with external environments felt unnatural, and incorporating execution results into reasoning was awkward. This barrier has largely fallen---with tool use and function calling emerging as core capabilities \citep{yao2022react, schick2023toolformer}, modern agents can execute code, interpret error messages, and refine their approach based on empirical evidence. The second barrier is more fundamental: the \textbf{Missing Oracle problem}. Executing code requires test cases with expected outputs, but for novel algorithmic problems, the correct output is precisely what we are trying to discover.

CodeSIM's authors confronted this oracle challenge directly. They explored augmenting test cases through self-consistency---generating tests twice and retaining only those appearing in both responses. The result was a \textbf{9.3\%p performance decline}, leading them to conclude that ``generating additional I/Os is a non-trivial task that may negatively impact final outcomes'' \citep{islam2025codesim}. However, we argue that the failure stemmed not from test generation itself, but from attempting to predict \textit{exact outputs} without ground truth. Our key insight is that verification need not require exact answers: by checking \textit{properties} rather than values (e.g., ``output length equals input length,'' ``result is a permutation of input''), we can judge correctness without an oracle. This property-based approach closes the verification loop that CodeSIM left open.

\paragraph{Two Dimensions of the Mental-Reality Gap.}
Our analysis reveals that this gap manifests along two orthogonal dimensions: the \textbf{Specification Gap} (overlooking edge cases during planning) and the \textbf{Verification Gap} (hallucinating execution traces during validation). These are independent---addressing one leaves the other open. Our ablation (Table~\ref{tab:ablation}) confirms: removing Shift-left Planning ([S]) causes a $-$23.7\%p drop, removing Live Execution ([L]) causes $-$7.9\%p, and the full system achieves performance that neither alone can reach.

Building on this insight, we propose \textbf{SolidCoder}, a multi-agent framework addressing \textit{both} dimensions. In the \textbf{S.O.L.I.D.} architecture: \textbf{S}hift-left Planning addresses the Specification Gap by forcing edge-case awareness before algorithm design; \textbf{L}ive Execution addresses the Verification Gap by replacing imagined traces with concrete runtime feedback; \textbf{O}racle-based Assertions enable verification without ground truth; \textbf{I}ntermediate Simulation provides a cost-effective pre-filter; and \textbf{D}efensive Accumulation prevents regression. Our contributions are:

\begin{itemize}
    \item We identify and characterize the \textbf{Mental-Reality Gap} as a \textit{two-dimensional} phenomenon: the \textit{Specification Gap} (edge-case blindness) and the \textit{Verification Gap} (execution hallucination)---a fundamental limitation in existing multi-agent code generation frameworks.
    \item We propose \textbf{SolidCoder}, a framework that jointly addresses \textit{both} dimensions through the \textbf{S.O.L.I.D.} architecture: Shift-left Planning for the Specification Gap, Live Execution for the Verification Gap, with Oracle assertions, Intermediate simulation, and Defensive accumulation as supporting mechanisms.
    \item We achieve new state-of-the-art pass@1 results with GPT-4o: \textbf{95.7\%} on HumanEval (+0.6\%p), \textbf{77.0\%} on CodeContests (+4.3\%p), and \textbf{26.7\%} on APPS (+3.4\%p). Our ablation demonstrates that both dimensions contribute substantially, with edge-case awareness providing the largest individual gain and execution grounding providing essential complementary value---together outperforming either component alone.
\end{itemize}

\section{Related Work}

\begin{figure*}[t]
    \centering
    \includegraphics[width=\textwidth]{fig2.png}
    \caption{\textbf{The Mental-Reality Gap in Action.} A comparative example on a list rotation problem. \textbf{Left (CodeSIM):} Mental Simulation traces through the code and incorrectly concludes ``Output matches expected. PASS''---the LLM hallucinates correct behavior despite a bug. \textbf{Right (SolidCoder):} Live Execution runs the actual code, revealing the bug through concrete failure: \texttt{AssertionError: [3,1,2] != [2,3,1]}. This demonstrates how grounding verification in reality catches bugs that mental simulation misses. (This example focuses on the Live Verification Loop; \textbf{[I]} Intermediate Simulation is depicted in Figure~\ref{fig:1}.)}
    \label{fig:2}
\end{figure*}

\subsection{Program Synthesis and LLMs for Code}
Program synthesis has a long-standing history in AI, evolving from early search-based methods \citep{manna1971toward, gulwani2011automating} to neural approaches \citep{feng2020codebert, ahmad2021unified}. The advent of Large Language Models (LLMs) has fundamentally transformed this landscape.

Codex \citep{chen2021evaluating} pioneered LLMs for code. GPT-4 \citep{achiam2023gpt} and GPT-4o \citep{hurst2024gpt} marked a turning point, with other frontier models \citep{team2023gemini, anthropic2024claude3} demonstrating comparable capabilities. Open-source alternatives \citep{roziere2023code, guo2024deepseek, hui2024qwen2} have rapidly closed the gap. More recently, reasoning models \citep{jaech2024openai, guo2025deepseek} have advanced code generation through techniques like RLVR \citep{wang2025reinforcement}, using execution feedback as training signal. As these models continue to improve, they increasingly serve as the ``brain'' for autonomous agents.

\subsection{Approaches to LLM Code Generation}
With capable LLMs as the foundation, the question becomes \textit{how} to best orchestrate them. Various prompting strategies have been developed to enhance code generation. Chain-of-Thought (CoT) \citep{wei2022chain} decomposes complex problems into intermediate reasoning steps. Retrieval-augmented generation \citep{parvez2021retrieval} leverages similar examples to guide generation. Self-planning \citep{jiang2024self} generates algorithmic plans before coding. Analogical reasoning \citep{yasunaga2023large} recalls and adapts solutions from related problems. ReAct \citep{yao2022react} interleaves reasoning with action execution. While these techniques improve single-pass generation, they remain limited by the LLM's inability to verify its own outputs through actual execution.

To address this limitation, recent research has shifted towards multi-agent frameworks that simulate the software development lifecycle. MapCoder \citep{islam2024mapcoder} introduced a pipeline of retrieval, planning, coding, and debugging agents, significantly outperforming single-agent baselines. CodeSIM \citep{islam2025codesim} further enhanced this by incorporating simulation-driven planning and debugging, allowing agents to verify their logic through step-by-step execution traces. Other approaches like Reflexion \citep{shinn2023reflexion} and LATS \citep{zhou2023language} leverage iterative self-correction and tree search, respectively. However, these frameworks still rely on the LLM's internal reasoning for verification, making them susceptible to hallucinated execution traces.

A separate line of work has explored leveraging actual code execution for debugging. LDB \citep{zhong2024debug} segments programs into basic blocks and tracks intermediate variable values during runtime, identifying discrepancies step-by-step. MGDebugger \citep{shi2024code} takes a hierarchical approach, decomposing code into a tree of sub-functions and debugging bottom-up. While these execution-based debuggers ground verification in reality, they operate as a \textit{second pass} after initial code generation and require ground-truth test cases to detect errors. SolidCoder addresses this limitation by performing verification \textit{in-loop} with property-based oracles rather than ground-truth outputs, enabling autonomous verification under missing-oracle constraints.

\section{SolidCoder Methodology}
\label{sec:methodology}

SolidCoder unifies the strengths of both paradigms while addressing their limitations. Like multi-agent frameworks, we operate within an integrated generation pipeline; like external debuggers, we leverage actual execution. Building upon the three-agent architecture of CodeSIM---comprising Planning, Coding, and Debugging agents---we introduce the \textbf{S.O.L.I.D.} framework to systematically address the Mental-Reality Gap. Our guiding principle is straightforward: \textit{don't imagine---execute}. Rather than relying on mental simulation for verification, we integrate concrete execution feedback directly into the generation loop.

Our key innovations are threefold: (1) we integrate live execution \textit{into} the generation loop rather than applying it post-hoc, (2) oracle-based assertions enable property verification without ground-truth answers, and (3) Defensive Accumulation ensures discovered bugs never regress. Table~\ref{tab:comparison} compares SolidCoder with existing code generation approaches.

\begin{table}[t]
    \centering
    \resizebox{\columnwidth}{!}{%
    \begin{tabular}{@{}lccccccc@{}}
        \toprule
        & \multicolumn{4}{c}{\textit{Generation}} & \multicolumn{3}{c}{\textit{Verification}} \\
        \cmidrule(lr){2-5} \cmidrule(l){6-8}
        \textbf{Approach} &
        \makecell{Exemplar\\Retrieval} &
        Plan &
        \makecell{Edge\\Aware} &
        \makecell{Mental\\Simulation} &
        \makecell{Live\\Execution} &
        Debug &
        \makecell{Defensive\\Accumulation} \\
        \midrule
        Reflexion     & \xmark & \xmark & \xmark & \xmark & \xmark & \cmark & \xmark \\
        Self-planning & \xmark & \cmark & \xmark & \xmark & \xmark & \xmark & \xmark \\
        Analogical    & \cmark & \xmark & \xmark & \xmark & \xmark & \xmark & \xmark \\
        LATS          & \xmark & \cmark & \xmark & \xmark & \xmark & \cmark & \xmark \\
        MapCoder      & \cmark & \cmark & \xmark & \xmark & \xmark & \cmark & \xmark \\
        CodeSIM       & \cmark & \cmark & \xmark & \cmark & \xmark & \cmark & \xmark \\
        \midrule
        \rowcolor{LightBlue}
        \textbf{SolidCoder} & \cmark & \cmark & \cmark & \cmark & \cmark & \cmark & \cmark \\
        \bottomrule
    \end{tabular}}
    \caption{Comparison of code generation approaches. SolidCoder integrates these capabilities within a single inference-time framework.}
    \label{tab:comparison}
\end{table}

\subsection{S.O.L.I.D. Components}

\paragraph{Shift-left Planning (S).} Traditional approaches address edge cases reactively during debugging. SolidCoder extends the Planning Agent to identify potential failure cases \textit{before} formulating the algorithmic plan. We prompt the LLM: \textit{``What worst-case inputs could break a naive solution?''} These edge cases (empty inputs, boundary values, corner cases) are then incorporated into the planning prompt, forcing robust algorithm design from the outset.

\paragraph{Oracle-based Assertions (O).} A fundamental challenge in test generation is the ``oracle problem''---determining correct outputs for novel inputs. Mental Simulation sidesteps this by having the LLM predict outputs, reintroducing hallucination risk. SolidCoder addresses this through \textit{property-based verification}: rather than checking exact outputs, we verify that results satisfy domain-invariant properties (e.g., a sorting function preserves length, maintains ordering, and produces a permutation). This transforms verification from ``Is this output correct?'' to ``Does this output satisfy required properties?''---answerable through execution.

\paragraph{Live Execution (L).} Rather than asking the LLM to imagine code behavior, we run it. Mental Simulation requires maintaining accurate state across potentially hundreds of execution steps---increasingly error-prone as complexity grows. Live Execution eliminates this by delegating state tracking to an actual interpreter. We execute generated code in a sandboxed environment (5-second timeout, restricted filesystem access), routing results to debugging on assertion failures or runtime errors. 

\paragraph{Intermediate Simulation (I).} A verified plan does not guarantee correct code---translation errors can occur. Immediately after code generation, we prompt the LLM to trace the code on a sample input. Why retain mental simulation if unreliable? The key difference is \textit{role}: in CodeSIM, it provides the terminal verdict; in SolidCoder, [I] is a preliminary filter followed by [L] Live Execution. If [I] misses a bug, [L] catches it through concrete execution. The critical guarantee: [L]---not [I]---provides the authoritative verdict.

\paragraph{Defensive Accumulation (D).} Iterative debugging introduces regression risk: fixing bug A may reintroduce bug B from a previous iteration. Defensive Accumulation maintains a persistent test suite that grows throughout debugging. Whenever Live Execution discovers a failing test case, that input-assertion pair joins an \texttt{accumulated\_tests} collection. Crucially, \textit{all} accumulated tests are re-executed after every code modification, providing a monotonicity guarantee: once code passes a test, regression triggers immediate detection.

\subsection{Integrated Workflow}
The five components integrate into the workflow detailed in Algorithm~\ref{alg:solidcoder} in Appendix~\ref{app:algorithm}. The Planning Agent applies \textbf{S} to identify edge cases before generating a plan. The Coding Agent translates the plan to code, followed by \textbf{I} to catch translation errors. The code enters the Live Verification loop: \textbf{O} generates property-based tests, \textbf{L} provides concrete execution feedback, and \textbf{D} accumulates failing tests. The algorithm terminates when code passes all sample I/Os and accumulated tests, or after exhausting $p=5$ planning and $d=5$ debugging iterations (following CodeSIM).

\section{Experiments}

\subsection{Datasets}
We adopt the evaluation protocol and problem subsets from CodeSIM \citep{islam2025codesim}, assessing generalization across three benchmarks that span a difficulty gradient:

\begin{itemize}[leftmargin=*, itemsep=2pt, topsep=2pt]
    \item \textbf{HumanEval} \citep{chen2021evaluating}: 164 function-level problems with short docstring specifications, often solvable in 1--2 lines (\textit{easy}).
    \item \textbf{CodeContests} \citep{li2022competition}: 165 Codeforces problems requiring algorithmic reasoning with precise I/O specifications and ${\sim}$200 hidden tests per problem (\textit{medium}).
    \item \textbf{APPS} \citep{hendrycks2021measuring}: 150 problems featuring complex rule-based tasks with ambiguous natural language descriptions (\textit{hard}).
\end{itemize}

We deliberately prioritize hypothesis fit over benchmark breadth. HumanEval, CodeContests, and APPS keep the task close to function-level or competitive-programming code generation, allowing us to observe verification failures while minimizing confounds from repository navigation, dependency resolution, and edit localization. Repository-level benchmarks such as SWE-bench \citep{jimenez2023swe} are important future validations, but they stress a partially different bottleneck---finding and coordinating edits across large codebases---than the missing-oracle verification problem studied here.

\subsection{Baselines and Metric}
We adopt the comparison set from CodeSIM \citep{islam2025codesim}, excluding Reflexion \citep{shinn2023reflexion} and LATS \citep{zhou2023language} because they are not reproducible under the publicly released CodeSIM codebase (no matching code for identical conditions). Included baselines are prompting methods (Direct, CoT \citep{wei2022chain}, Self-Planning \citep{jiang2024self}), retrieval-augmented approaches (Analogical \citep{yasunaga2023large}), and multi-agent frameworks: MapCoder \citep{islam2024mapcoder} and CodeSIM \citep{islam2025codesim}. We report \textbf{pass@1}: a problem is solved if the first generated solution passes all hidden test cases.

\subsection{Experimental Setup}
\label{sec:experimental_setup}
We evaluate on three models: \textbf{GPT-4o} (\texttt{gpt-4o-2024-08-06}) as our primary baseline for direct comparison with prior work \citep{islam2024mapcoder, islam2025codesim}; \textbf{GPT-OSS-120B} \citep{agarwal2025gpt}, an open-source GPT model post-trained with RLVR \citep{wang2025reinforcement}; and \textbf{Grok-4.1-Fast} \citep{xai2025grok4card}, a non-GPT frontier model also post-trained with RL, with strong math and coding performance. This selection tests generalization across three axes: proprietary vs.\ open-source, pre- vs.\ post-RLVR, and GPT vs.\ non-GPT architectures.

For GPT-4o, we use the OpenAI API\footnote{\url{https://platform.openai.com/docs/models/gpt-4o}} with temperature $\tau{=}0$ and top-$p{=}0.95$, adopting HumanEval baselines from the CodeSIM paper and reproducing CodeContests/APPS with the official codebase.\footnote{\url{https://github.com/kagnlp/CodeGenerator}} The additional models use default configurations from their OpenRouter\footnote{\url{https://openrouter.ai}} API pages. All experiments follow CodeSIM hyperparameters ($p{=}5$ planning iterations, $d{=}5$ debugging iterations); SolidCoder uses $a{=}3$ for assumption breaking.

\section{Results}

\subsection{Main Results}

\begin{table*}[t]
    \centering
    \small
    \renewcommand{\arraystretch}{0.92}
    \setlength{\tabcolsep}{4pt}
    \resizebox{\textwidth}{!}{%
    \begin{tabular}{@{}l|ccccc|cc@{}}
        \toprule
        \textbf{Model} & \textbf{Direct} & \textbf{CoT} & \textbf{SelfPlanning} & \textbf{Analogical} & \textbf{MapCoder} & \textbf{CodeSIM} & \textbf{SolidCoder} \\
        \midrule
        \rowcolor{green!15}
        \multicolumn{8}{c}{\textbf{HumanEval}~\citep{chen2021evaluating} {\small\textcolor{green!60!black}{(EASY)}}} \\
        GPT-4o$^*$ & 90.2 & 90.9 & 89.0 & 88.4 & 90.2 & 95.1 & \textbf{95.7} \\
        GPT-OSS-120B & 69.5 & 96.3 & 90.8 & 89.0 & 61.0 & \textbf{98.2} & \textbf{98.2} \\
        Grok-4.1-Fast & 88.4 & 96.9 & 96.9 & 95.7 & 95.7 & \textbf{97.6} & \textbf{97.6} \\
        \cmidrule{1-8}
        \textit{Average} & \textit{82.7} & \textit{94.7} & \textit{92.2} & \textit{91.0} & \textit{82.3} & \textit{97.0} & \textit{\textbf{97.2}} \\
        \midrule
        \rowcolor{yellow!20}
        \multicolumn{8}{c}{\textbf{CodeContests}~\citep{li2022competition} {\small\textcolor{orange!80!black}{(MEDIUM)}}} \\
        GPT-4o$^*$ & 42.4 & 44.2 & 49.1 & 30.3 & 69.1 & 72.7 & \textbf{77.0} \\
        GPT-OSS-120B & 75.8 & 75.2 & 75.2 & 75.2 & 44.8 & 87.9 & \textbf{92.1} \\
        Grok-4.1-Fast & 79.4 & 85.4 & 81.2 & 77.0 & 83.6 & 95.2 & \textbf{98.2} \\
        \cmidrule{1-8}
        \textit{Average} & \textit{65.9} & \textit{68.3} & \textit{68.5} & \textit{60.8} & \textit{65.8} & \textit{85.3} & \textit{\textbf{89.1}} \\
        \midrule
        \rowcolor{red!15}
        \multicolumn{8}{c}{\textbf{APPS}~\citep{hendrycks2021measuring} {\small\textcolor{red!70!black}{(HARD)}}} \\
        GPT-4o$^*$ & 10.7 & 17.3 & 14.7 & 14.0 & 20.7 & 23.3 & \textbf{26.7} \\
        GPT-OSS-120B & 35.3 & 32.7 & 34.7 & 30.7 & 24.0 & 39.3 & \textbf{40.7} \\
        Grok-4.1-Fast & 37.3 & 34.7 & 37.3 & 36.0 & 33.3 & 41.3 & \textbf{42.0} \\
        \cmidrule{1-8}
        \textit{Average} & \textit{27.8} & \textit{28.2} & \textit{28.9} & \textit{26.9} & \textit{26.0} & \textit{34.6} & \textit{\textbf{36.5}} \\
        \bottomrule
    \end{tabular}}
    \caption{Pass@1 (\%) across seven strategies and three models. Benchmarks are ordered by difficulty. $^*$GPT-4o enables comparison with prior work; the remaining models are RL post-trained. SolidCoder consistently outperforms CodeSIM, with the largest gain on CodeContests (+3.8\%p avg). \textbf{Bold}: best per row.}
    \label{tab:main_and_multimodel}
\end{table*}

Table~\ref{tab:main_and_multimodel} presents results across three benchmarks. SolidCoder matches or outperforms CodeSIM in all nine model-benchmark combinations. On GPT-4o, SolidCoder achieves \textbf{95.7\%} on HumanEval (+0.6\%p), \textbf{77.0\%} on CodeContests (+4.3\%p), and \textbf{26.7\%} on APPS (+3.4\%p)---enabling direct comparison with the CodeSIM paper~\citep{islam2025codesim}.

The pattern of improvements reveals a systematic relationship with benchmark difficulty. HumanEval, where current LLMs already exceed 97\% average performance, shows minimal gains---the benchmark is approaching saturation, leaving little room for verification-based improvements. CodeContests, by contrast, exhibits the largest and most consistent gains across all models: GPT-4o +4.3\%p, GPT-OSS-120B +4.2\%p, and Grok-4.1-Fast +3.0\%p (averaging +3.8\%p). We attribute this to CodeContests occupying a ``sweet spot'' in difficulty---problems are complex enough that mental simulation becomes unreliable, yet remain tractable enough for execution-grounded verification to effectively catch and correct errors. Notably, Grok-4.1-Fast with SolidCoder reaches \textbf{98.2\%}, approaching ceiling performance on this competitive programming benchmark.

On APPS, improvements vary by model: GPT-4o gains +3.4\%p, while RL post-trained models show smaller improvements (GPT-OSS-120B +1.4\%p, Grok-4.1-Fast +0.7\%p). APPS poses such extreme difficulty that the bottleneck shifts from \textit{verification} to \textit{generation}---even perfect verification cannot salvage fundamentally flawed algorithms. However, the same RL models still show substantial gains on CodeContests (+4.2\%p and +3.0\%p), indicating that while RLVR improves code generation, models still rely on mental simulation when evaluating their outputs at inference time. SolidCoder's live execution provides value by replacing this unreliable self-assessment with concrete feedback. Overall, these results validate that the Mental-Reality Gap is a fundamental limitation of mental simulation itself, and execution-grounded verification provides consistent improvements across model families and training paradigms.

\subsection{Ablation Study}

To quantify the contribution of each S.O.L.I.D. component, we conduct a leave-one-out ablation study on CodeContests using GPT-4o, our primary model for comparison with prior work. We select this benchmark because it offers an appropriate difficulty level where component contributions are likely to be clearly distinguishable. On HumanEval, current LLMs approach saturation (95\%+), which would make subtle variations difficult to observe reliably. Conversely, APPS poses such high difficulty that component differences may be obscured by uniformly low scores. CodeContests strikes a balance, providing sufficient headroom to reveal each component's contribution.

Each ablation removes exactly one component from the full pipeline. In \textit{w/o [L] Live Execution}, the concrete sandbox execution stage is replaced by mental assumption-breaking, where the model reasons about correctness through text-based tracing rather than actual execution; this isolates the value of execution grounding. In \textit{w/o [O] Oracle-based Assertions}, live execution is retained but tests rely on crash detection rather than model-generated property assertions; this isolates the contribution of oracle-style specifications.

\begin{table}[t]
    \centering
    \small
    \begin{tabular}{@{}lcc@{}}
        \toprule
        \textbf{Configuration} & \textbf{Pass@1} & \textbf{$\Delta$} \\
        \midrule
        GPT-4o Direct  & 42.4 & \textcolor{CrossRed}{$-$34.6} \\
        \addlinespace[3pt]
        \quad w/o \textbf{S}hift-left Planning & 53.3 & \textcolor{CrossRed}{$-$23.7} \\
        \quad w/o \textbf{I}ntermediate Simulation & 64.0 & \textcolor{CrossRed}{$-$13.0} \\
        \quad w/o \textbf{O}racle-based Assertions & 65.4 & \textcolor{CrossRed}{$-$11.6} \\
        \quad w/o \textbf{L}ive Execution & 69.1 & \textcolor{CrossRed}{$-$7.9} \\
        \quad w/o \textbf{D}efensive Accumulation & 70.3 & \textcolor{CrossRed}{$-$6.7} \\
        \midrule
        \textbf{GPT-4o SolidCoder (Full)} & \textbf{77.0} & \textcolor{gray}{--} \\
        \bottomrule
    \end{tabular}
    \caption{Ablation on CodeContests (GPT-4o). $\Delta$ indicates drop from Full. Components ranked by impact.}
    \label{tab:ablation}
\end{table}

\begin{table*}[t]
    \centering
    \small
    \setlength{\tabcolsep}{3pt}
    \begin{tabular}{@{}lccccccccc@{}}
        \toprule
        & \multicolumn{3}{c}{\textbf{GPT-4o}$^*$} & \multicolumn{3}{c}{\textbf{GPT-OSS-120B}} & \multicolumn{3}{c}{\textbf{Grok-4.1-Fast}} \\
        \cmidrule(lr){2-4} \cmidrule(lr){5-7} \cmidrule(lr){8-10}
        & MapCoder & CodeSIM & SolidCoder & MapCoder & CodeSIM & SolidCoder & MapCoder & CodeSIM & SolidCoder \\
        \midrule
        \rowcolor{green!10}
        \multicolumn{10}{c}{\textbf{HumanEval}~\citep{chen2021evaluating} \textcolor{green!60!black}{\scriptsize(EASY)}} \\
        Pass@1 (\%) & 90.2 & 95.1 & \textbf{95.7} & 61.0 & \textbf{98.2} & \textbf{98.2} & 95.7 & 97.0 & \textbf{97.6} \\
        API Calls & 9 & \underline{4} & 8 & 6 & \underline{3} & 19 & 8 & \underline{4} & 12 \\
        Tokens (K) & 6.6 & \underline{3.8} & 7.5 & 10.0 & \underline{5.6} & 40.9 & 18.7 & \underline{14.3} & 47.8 \\
        \addlinespace[3pt]
        \rowcolor{yellow!15}
        \multicolumn{10}{c}{\textbf{CodeContests}~\citep{li2022competition} \textcolor{yellow!60!black}{\scriptsize(MEDIUM)}} \\
        Pass@1 (\%) & 69.1 & 72.7 & \textbf{77.0} & 44.8 & 87.9 & \textbf{92.1} & 83.6 & 95.2 & \textbf{98.2} \\
        API Calls & 11 & \underline{11} & 16 & \underline{4} & 6 & 15 & 8 & \underline{4} & 10 \\
        Tokens (K) & 20.9 & \underline{20.9} & 31.9 & \underline{11.5} & 28.3 & 58.3 & 44.2 & \underline{35.0} & 96.7 \\
        \addlinespace[3pt]
        \rowcolor{red!10}
        \multicolumn{10}{c}{\textbf{APPS}~\citep{hendrycks2021measuring} \textcolor{red!60!black}{\scriptsize(HARD)}} \\
        Pass@1 (\%) & 20.7 & 23.3 & \textbf{26.7} & 24.0 & 39.3 & \textbf{40.7} & 33.3 & 41.3 & \textbf{42.0} \\
        API Calls & \underline{19} & 26 & 35 & \underline{9} & 21 & 48 & 15 & \underline{20} & 47 \\
        Tokens (K) & \underline{36.4} & 49.3 & 60.4 & \underline{31.3} & 73.3 & 177.4 & 97.5 & \underline{266.5} & 520.9 \\
        \bottomrule
    \end{tabular}
    \caption{Efficiency comparison across models. \underline{Underline}: most efficient; \textbf{bold}: highest Pass@1 per model-benchmark. GPT-OSS-120B shows degraded MapCoder performance due to XML parsing incompatibility. $^*$Primary model for comparison with prior work.}
    \label{tab:efficiency}
\end{table*}

Table~\ref{tab:ablation} presents the results. The findings validate our two-dimensional analysis of the Mental-Reality Gap introduced in Section~\ref{sec:intro}.

\paragraph{Specification Gap Dominates on CodeContests.}
The largest drops come from specification-oriented components: removing [S] Shift-left Planning causes $-$23.7\%p, and removing [I] Intermediate Simulation causes $-$13.0\%p. Together, these account for the majority of SolidCoder's improvement over the Direct baseline. This confirms that for algorithmic competition problems, \textit{edge-case blindness} is the predominant failure mode---models generate solutions that work on typical inputs but fail on boundary conditions.

\paragraph{Verification Gap Remains Critical.}
Despite smaller absolute contribution, [L] Live Execution's $-$7.9\%p drop represents cases where mental simulation would have \textit{incorrectly validated buggy code}. These are precisely the ``Mental-Reality Gap'' failures that motivated our work: the model traces through incorrect logic and hallucinates a correct result. Without [L], such errors escape detection entirely. The 7.9\%p improvement may appear modest relative to [S], but it reflects a categorically different error type that specification improvements cannot address.

\paragraph{Component Synergy.}
The full system (77.0\%) substantially outperforms any single-component removal, indicating synergistic interaction. Edge-case identification ([S]) generates more comprehensive test scenarios, which live execution ([L]) then validates with concrete feedback. Neither component alone achieves comparable results: [S] without [L] still relies on hallucination-prone mental verification; [L] without [S] only tests the happy path that the model already handles well. The near-additivity of component contributions supports this interpretation---both components are essential for addressing both gaps.

\paragraph{Supporting Components.}
[O] Oracle-based Assertions ($-$11.6\%p) enables property verification without ground truth---a prerequisite for [L] to function on novel inputs. [D] Defensive Accumulation ($-$6.7\%p) prevents regression during iterative debugging, ensuring that discovered bugs stay fixed.

\subsection{Efficiency Analysis}

SolidCoder's execution-grounded verification introduces additional cost. Table~\ref{tab:efficiency} compares efficiency across models.

On easy problems like HumanEval, SolidCoder's overhead is inefficient: GPT-4o uses +97\% more tokens for only +0.6\%p gain. However, on problems where base LLMs struggle---CodeContests and APPS---the additional cost yields proportional improvements: +50\% tokens for +4.3\%p on CodeContests, +23\% for +3.4\%p on APPS. RL post-trained models show similar patterns, achieving consistent gains on CodeContests (+4.2\%p for GPT-OSS, +3.0\%p for Grok) despite higher token overhead.

This suggests that difficulty-aware routing could maximize SolidCoder's value: applying lightweight paths to simple problems while reserving the full verification loop for complex tasks where mental simulation is most likely to fail. In this sense, SolidCoder can be viewed not only as a fixed pipeline but also as an inference-time framework whose verification intensity can be adapted to task difficulty.

\subsection{Qualitative Example}
Figure~\ref{fig:2} concretizes the Mental-Reality Gap on a list rotation problem: mental simulation hallucinates correct behavior for buggy code, whereas SolidCoder's Live Execution immediately exposes the fault through a concrete \texttt{AssertionError}. A complete end-to-end trace of the full S.O.L.I.D. pipeline, including all LLM inputs and outputs, is provided on a separate problem in Appendix~\ref{app:example}.

\section{Conclusion and Future Work}
We proposed \textbf{SolidCoder} with the \textbf{S.O.L.I.D.} architecture to bridge the Mental-Reality Gap through concrete execution, guided by a simple principle: \textit{don't imagine---execute}. Our analysis showed that this gap manifests in two orthogonal dimensions---the Specification Gap and the Verification Gap---and our ablation indicates that [S] reduces the dominant failure mode in algorithmic tasks, [O] enables verification without ground-truth outputs, and [L] catches errors that specification improvement alone cannot remove. With GPT-4o, SolidCoder achieves \textbf{95.7\%} on HumanEval, \textbf{77.0\%} on CodeContests, and \textbf{26.7\%} on APPS; these gains generalize across RL post-trained models and are most pronounced on medium-difficulty tasks where mental simulation is unreliable yet problems remain tractable. More broadly, our findings suggest that the next bottleneck in code generation lies not only in stronger generation but in stronger self-evaluation. Even though our experiments focus on function-level benchmarks, the underlying principle generalizes: as models become more capable generators, it matters increasingly how they falsify and revise their own hypotheses. Future work will explore difficulty-aware routing and more structural ways to reduce self-verification errors, and we release our code and framework to support that line of research.

\section*{Limitations}

Our work has several limitations that point to future research directions.

\textbf{First}, \textbf{evaluation scope}. Live Execution currently supports Python only; extending to other languages requires language-specific sandboxing---a tractable engineering effort. Additionally, our evaluation focuses on function-level benchmarks that cleanly isolate the Mental-Reality Gap. Generalization to repository-level tasks (e.g., SWE-bench \citep{jimenez2023swe}) remains to be validated, though S.O.L.I.D. components could complement existing repository-level frameworks.

\textbf{Second}, \textbf{LLM dependency}. While Oracle-based Assertions reduce reliance on exact output prediction, hallucinated properties remain possible, and [O] is most effective when correctness can be partially expressed through structural constraints or invariants rather than near-exact outputs. Moreover, when the same LLM generates code, properties, and validates tests, systematic biases may propagate. [L] partially mitigates this via model-independent feedback, but heterogeneous model combinations are an important future direction.

\textbf{Third}, \textbf{experimental scope}. We evaluate three models (GPT-4o, GPT-OSS-120B, Grok-4.1-Fast) but did not test Gemini or Claude due to cost constraints, and benchmark contamination cannot be ruled out---though identical evaluation conditions preserve relative comparisons. SolidCoder also consumes more tokens than CodeSIM (+50\% on CodeContests, +97\% on HumanEval), and our ablation covers only CodeContests with GPT-4o; compute-matched comparisons and cross-benchmark ablations remain for future work.

\section*{Ethical Considerations}

Our work aims to improve the reliability of autonomous code generation, which inherently carries dual-use implications. While \textbf{SolidCoder} enhances developer productivity and debugging capabilities, we acknowledge that the same mechanisms could potentially be misused to generate malicious software with higher success rates.

To mitigate operational risks associated with \textbf{Live Execution} of untrusted model-generated code, we implemented sandboxing measures in our experiments, including network blocking and execution timeouts. We strongly emphasize that any real-world deployment of execution-based agents must incorporate robust, OS-level containment (e.g., containers or microVMs) to prevent sandbox escapes or resource exhaustion.

Furthermore, our framework relies on large language models, and thus generated code or test cases may reflect biases present in the training data. Users should treat the system's outputs as suggestions requiring human review, particularly in safety-critical applications. Finally, we utilized standard public benchmarks (HumanEval, CodeContests, APPS) that do not contain personally identifiable information.

\section*{Acknowledgements}
This work was supported by Institute of Information and Communications Technology Planning and Evaluation(IITP) grant funded by the Korea government(MSIT) (RS-2019-II190004, Development of semi-supervised learning language intelligence technology and Korean tutoring service for foreigners).

\bibliography{custom}

\appendix

\section{Algorithm Details}
\label{app:algorithm}

Algorithm~\ref{alg:solidcoder} presents the complete SolidCoder workflow. The algorithm consists of three main stages: (1) edge case identification and plan generation via Shift-left Planning, (2) verification through Oracle tests and Live Execution in the Assumption Breaking Phase, and (3) iterative code refinement in the Traditional Debugging Phase. Each S.O.L.I.D. component is marked in \textcolor{blue}{blue}.

\begin{algorithm}[h]
\caption{The SolidCoder Workflow}
\label{alg:solidcoder}
\small
\begin{algorithmic}[1]
\Require Problem $P$, Sample I/O $T$, max\_plan $p$, max\_debug $d$, max\_assume $a$
\Ensure Solution Code $S$
\State $accumulated \leftarrow \emptyset$
\For{$i = 1$ to $p$}
    \State \textcolor{blue}{\textbf{[S]}} $edges \leftarrow \Call{IdentifyEdgeCases}{P}$
    \State $plan \leftarrow \Call{GeneratePlan}{P, edges}$
    \State $valid \leftarrow \Call{SimulatePlan}{P, plan}$ \Comment{Mental Verification}
    \If{$\neg valid$}
        \State $plan \leftarrow \Call{RefinePlan}{P, plan}$
    \EndIf
    \State $S \leftarrow \Call{GenerateCode}{P, plan}$
    \State \textcolor{blue}{\textbf{[I]}} $S \leftarrow \Call{IntermediateSimulation}{P, S}$ \Comment{Complementary filter}

    \For{$k = 1$ to $a$} \Comment{Assumption Breaking Phase}
        \State \textcolor{blue}{\textbf{[O]}} $test \leftarrow \Call{GenerateOracleTest}{P, S}$
        \State \textcolor{blue}{\textbf{[L]}} $result \leftarrow \Call{LiveExecute}{S, test}$
        \If{$result = \texttt{FAIL}$}
            \State \textcolor{blue}{\textbf{[D]}} $accumulated \leftarrow accumulated \cup \{test\}$
            \State $S \leftarrow \Call{FixCode}{P, S, test}$
        \Else
            \State \textbf{break} \Comment{Verification Passed}
        \EndIf
    \EndFor

    \If{$\Call{PassAll}{S, T \cup accumulated}$}
        \State \Return $S$
    \EndIf

    \For{$j = 1$ to $d$} \Comment{Traditional Debugging Phase}
        \State $S \leftarrow \Call{DebugCode}{P, plan, S}$
        \If{$\Call{PassAll}{S, T \cup accumulated}$}
            \State \Return $S$
        \EndIf
    \EndFor
\EndFor
\State \Return $S$
\end{algorithmic}
\end{algorithm}

\section{S.O.L.I.D. Component Prompts}
\label{app:prompts}
This appendix presents the complete prompts used in SolidCoder's S.O.L.I.D. components. Variables in curly braces (e.g., \texttt{\{problem\}}) are replaced with actual values at runtime.

\subsection{[S] Shift-left Planning Prompt}
The Shift-left prompt extends standard planning by requiring edge case identification \textit{before} algorithm design:

\begin{promptbox}[title={\textcolor{white}{[S] Shift-left Planning Prompt}}]{ShiftColor}
\small\ttfamily
You are an expert software engineer.\\
You are given a problem and you need to generate a detailed plan to solve it.\\[0.5em]
\#\# Problem\\
\{problem\}\\[0.5em]
\#\# Killer Edge Cases\\
Identify 3 potential edge cases that could break a naive solution:\\
1. Empty/minimal input\\
2. Maximum constraint input\\
3. Special pattern (all same, alternating, etc.) or Boundary values\\[0.5em]
\#\# Plan\\
Write a detailed plan that handles these edge cases.\\
The plan should be step-by-step and easy to implement.
\end{promptbox}

\subsection{[O] Oracle-based Assertion Prompt}
The Oracle prompt instructs the LLM to act as a ``Red Team'' tester, generating property-based assertions rather than exact output predictions:

\begin{promptbox}[title={\textcolor{white}{[O] Oracle-based Assertion Prompt}}]{OracleColor}
\small\ttfamily
You are an expert software tester (Red Team).\\
Your goal is to break the following code by finding hidden assumptions.\\[0.5em]
\#\# Problem\\
\{problem\}\\[0.5em]
\#\# Code\\
```\{language\}\\
\{code\}\\
```\\[0.5em]
\#\# Your Task\\
1. Identify a weak assumption (Type, Value, Structure, or Relationship).\\
2. Produce a **Python test script** that calls the target function with a breaking input.\\
3. Include an assert that would fail if the assumption is violated.\\[0.5em]
Format:\\
Assumption: <short text>\\
Test Script:\\
```python\\
\# call the function defined above\\
result = <call>\\
assert <oracle about result>\\
```
\end{promptbox}

\subsection{[L] Live Execution}
Live Execution does not use a separate prompt---it directly executes the test scripts generated by the Oracle component in a sandboxed Python environment. The execution wrapper is shown below:

\begin{codebox}[title={\textcolor{white}{[L] Live Execution Wrapper}}, colframe=LiveColor!80!black, colback=LiveColor!5!white, boxed title style={colback=LiveColor!80!black}]
\small\ttfamily
def \_concrete\_verify\_script(self, code, test\_script, timeout=5):\\
\hspace*{1em}full\_code = f\textquotedbl\textquotedbl\textquotedbl\\
\hspace*{1em}import sys\\
\hspace*{1em}import math\\
\hspace*{1em}from typing import List, Dict, Any, Optional, Union, Tuple\\[0.3em]
\hspace*{1em}\{code\}\\[0.3em]
\hspace*{1em}\{test\_script\}\\
\hspace*{1em}\textquotedbl\textquotedbl\textquotedbl\\
\hspace*{1em}try:\\
\hspace*{2em}safe\_builtins = builtins.\_\_dict\_\_.copy()\\
\hspace*{2em}safe\_builtins["input"] = lambda *\_, **\_\_: $\backslash$\\
\hspace*{3em}(\_ for \_ in ()).throw(RuntimeError("input() disabled"))\\
\hspace*{2em}exec\_globals = \{"\_\_builtins\_\_": safe\_builtins\}\\
\hspace*{2em}function\_with\_timeout(exec, (full\_code, exec\_globals), timeout)\\
\hspace*{2em}return "PASS"\\
\hspace*{1em}except AssertionError:\\
\hspace*{2em}return "FAIL\_ASSERT"\\
\hspace*{1em}except Exception:\\
\hspace*{2em}return "FAIL\_CRASH"
\end{codebox}

\noindent See Appendix~\ref{app:safety} for detailed safety mechanisms.

\subsection{[I] Intermediate Simulation Prompt}
The Intermediate Simulation prompt asks the LLM to mentally trace code execution before live verification:

\begin{promptbox}[title={\textcolor{white}{[I] Intermediate Simulation Prompt}}]{InterColor}
\small\ttfamily
\#\# Problem\\
\{problem\}\\[0.5em]
\#\# Code\\
```\{language\}\\
\{code\}\\
```\\[0.5em]
\#\# Your Task: Mental Execution\\
1. Select a sample input.\\
2. Trace the code execution step-by-step.\\
3. Track variable values.\\
4. Predict the final output.\\[0.5em]
If you find a logic error or mismatch with the plan, output: **CODE\_SIMULATION\_FAILED**\\
If the code seems correct, output: **CODE\_SIMULATION\_PASSED**
\end{promptbox}

\subsection{[D] Defensive Accumulation}
Defensive Accumulation is implemented as a runtime mechanism rather than a prompt. When Live Execution detects a failing test case, the test is automatically added to an \texttt{accumulated\_tests} collection. All subsequent code modifications must pass \textit{every} accumulated test before being accepted:

\begin{codebox}[title={\textcolor{white}{[D] Defensive Accumulation Logic}}, colframe=DefensiveColor!80!black, colback=DefensiveColor!5!white, boxed title style={colback=DefensiveColor!80!black}]
\small\ttfamily
\textcolor{gray}{\# In the assumption breaking loop:}\\
verify\_status = self.\_concrete\_verify\_script(code, test\_script)\\[0.3em]
if verify\_status in \{"FAIL\_CRASH", "FAIL\_ASSERT"\}:\\
\hspace*{1em}\textcolor{gray}{\# [D] Defensive Accumulation: store failing test}\\
\hspace*{1em}if self.enable\_defensive\_test:\\
\hspace*{2em}self.accumulated\_inputs.append(test\_script)\\
\hspace*{2em}additional\_io.append(test\_script)\\[0.3em]
\hspace*{1em}\textcolor{gray}{\# Fix the code based on the failing test}\\
\hspace*{1em}code = self.\_fix\_with\_test\_script(problem, code, test\_script)\\
\hspace*{1em}continue \textcolor{gray}{\# Re-verify with accumulated tests}
\end{codebox}

\subsection{Test Case Validation (Judge Prompt)}
To prevent false positives from invalid test cases, SolidCoder includes a Judge component that validates generated tests:

\begin{promptbox}[title={\textcolor{white}{Judge Prompt}}, colframe=gray!70!black, colback=gray!5!white, boxed title style={colback=gray!70!black}]{gray}
\small\ttfamily
You are an impartial Judge.\\[0.5em]
\#\# Problem\\
\{problem\}\\[0.5em]
\#\# Proposed Test Case\\
```python\\
\{test\_script\}\\
```\\[0.5em]
\#\# Task\\
Determine if this test case is VALID and CORRECT for the given problem.\\
1. Does the input satisfy all constraints (e.g. range, type, format)?\\
2. Is the asserted output (if any) logically correct?\\
3. Is it a fair test?\\[0.5em]
If the test case is valid, output: **VALID**\\
If it violates constraints or expects wrong output, output: **INVALID**
\end{promptbox}

\subsection{Code Fix Prompt}
When a vulnerability is found, SolidCoder uses this prompt to fix the code:

\begin{promptbox}[title={\textcolor{white}{Code Fix Prompt}}, colframe=gray!70!black, colback=gray!5!white, boxed title style={colback=gray!70!black}]{gray}
\small\ttfamily
\#\# Problem\\
\{problem\}\\[0.5em]
\#\# Code\\
```\{language\}\\
\{code\}\\
```\\[0.5em]
\#\# Vulnerability Found\\
The following test script failed (Crash or Assertion Error):\\
```python\\
\{test\_script\}\\
```\\[0.5em]
\#\# Task\\
Fix the code to handle this edge case.\\
Wrap the code in ```\{language\} ... ``` block.
\end{promptbox}

\section{Live Execution Safety Mechanisms}
\label{app:safety}
SolidCoder implements several safety mechanisms for Live Execution to prevent malicious or buggy code from affecting the host system. Table~\ref{tab:safety} summarizes these mechanisms. Full implementation details are provided in the \textcolor[rgb]{0.6, 0.6, 0.0}{[L] Live Execution} code listing in Appendix~\ref{app:prompts}.

\begin{table*}[t]
\centering
\small
\renewcommand{\arraystretch}{1.3}
\begin{tabular}{@{}>{\raggedright\arraybackslash}p{2.8cm}>{\raggedright\arraybackslash}p{4.2cm}>{\raggedright\arraybackslash}p{8.5cm}@{}}
\toprule
\rowcolor{SafetyColor!15}
\textbf{Mechanism} & \textbf{Implementation} & \textbf{Purpose \& Rationale} \\
\midrule
Timeout & 5-second limit per test via \texttt{function\_with\_timeout} & Prevents infinite loops from blocking the pipeline. Correct solutions complete in milliseconds; 5s provides margin while catching runaway executions. \\
\arrayrulecolor{gray!30}\hline\arrayrulecolor{black}
Network Blocking & \texttt{block\_network=True} (default) & Prevents generated code from making outbound network calls, eliminating data exfiltration and external dependency risks during sandboxed execution. \\
\arrayrulecolor{gray!30}\hline\arrayrulecolor{black}
Input Blocking & \texttt{input()} $\rightarrow$ \texttt{RuntimeError} stub & Prevents scripts from hanging indefinitely waiting for user input that will never arrive in automated execution. \\
\arrayrulecolor{gray!30}\hline\arrayrulecolor{black}
Isolated Namespace & Fresh \texttt{exec\_globals} per execution & Prevents test scripts from interfering with each other or accumulating state. Each test starts with a clean environment. \\
\arrayrulecolor{gray!30}\hline\arrayrulecolor{black}
Exception Classification & \texttt{PASS}, \texttt{FAIL\_ASSERT}, \texttt{FAIL\_CRASH} & Distinguishes logical failures (\texttt{AssertionError}) from runtime bugs. Both trigger Defensive Accumulation with different diagnostic signals. \\
\bottomrule
\end{tabular}
\caption{Safety mechanisms implemented in SolidCoder's Live Execution component to ensure reliable and secure code evaluation.}
\label{tab:safety}
\end{table*}

\onecolumn

\section{Complete Example}
\label{app:example}

This section presents a complete execution trace of SolidCoder solving HumanEval problem \#0 (\texttt{has\_close\_elements}). The trace demonstrates the full S.O.L.I.D. pipeline: \textbf{[S]} Shift-left Planning identifies edge cases before generating the plan; \textbf{[I]} Intermediate Simulation traces the generated code mentally; \textbf{[O]} Oracle-based Assumption Breaking generates adversarial test cases; and \textbf{[L]} Live Execution validates correctness through actual runtime. In this example, the code passes on the first attempt, so \textbf{[D]} Defensive Accumulation is not triggered. All prompts and LLM responses are shown verbatim from the experiment log.

\vspace{3mm}

\begin{tcolorbox}[breakable, colback=white, colframe=black!70, boxrule=0.5pt, arc=2pt, left=4mm, right=4mm, top=3mm, bottom=3mm]

\vspace{2mm}
\textbf{\large Complete Execution Trace: HumanEval/0 (\texttt{has\_close\_elements}) with GPT-4o}
\hrule
\vspace{5mm}

\noindent\textbf{\Large Problem}

\begin{verbatim}
from typing import List

def has_close_elements(numbers: List[float], threshold: float) -> bool:
    """ Check if in given list of numbers, are any two numbers closer to each 
    other than given threshold.
    >>> has_close_elements([1.0, 2.0, 3.0], 0.5)
    False
    >>> has_close_elements([1.0, 2.8, 3.0, 4.0, 5.0, 2.0], 0.3)
    True
    """
\end{verbatim}

\vspace{0.5cm} \hrule \vspace{0.5cm}

\noindent\textbf{\Large [S] Shift-left Planning}

\textbf{Prompt:}
\begin{verbatim}
You are an expert software engineer.
You are given a problem and you need to generate a detailed plan to solve it.

## Problem
[problem code shown above]

## Killer Edge Cases
Identify 3 potential edge cases that could break a naive solution:
1. Empty/minimal input
2. Maximum constraint input
3. Special pattern (all same, alternating, etc.) or Boundary values

## Plan
Write a detailed plan that handles these edge cases.
The plan should be step-by-step and easy to implement.
\end{verbatim}

\textbf{LLM Response:}

To solve the problem of determining if any two numbers in a list are closer to each other than a given threshold, we need to consider potential edge cases and develop a robust plan. Here's a detailed plan to address the problem and handle edge cases:

\textbf{\#\#\# Edge Cases}

\textbf{1. Empty/Minimal Input:}
\begin{itemize}
    \item An empty list or a list with a single element should return \texttt{False} because there are no pairs of numbers to compare.
\end{itemize}

\textbf{2. Maximum Constraint Input:}
\begin{itemize}
    \item Consider the scenario where the list contains the maximum number of elements allowed by the system's memory. The solution should be efficient in terms of time and space complexity to handle large inputs.
\end{itemize}

\textbf{3. Special Pattern or Boundary Values:}
\begin{itemize}
    \item Lists where all elements are the same, which should return \texttt{True} if the threshold is greater than zero.
    \item Lists with alternating values or values that are very close to each other but not within the threshold.
    \item Lists with very large or very small floating-point numbers to test precision and handling of floating-point arithmetic.
\end{itemize}

\textbf{\#\#\# Plan}

\textbf{1. Initial Checks:}
\begin{itemize}
    \item If the list \texttt{numbers} is empty or contains only one element, return \texttt{False} immediately since no pairs exist.
\end{itemize}

\textbf{2. Sorting:}
\begin{itemize}
    \item Sort the list of numbers. Sorting helps in efficiently finding close elements because if two numbers are close, they will be adjacent in a sorted list.
\end{itemize}

\textbf{3. Iterate and Compare:}
\begin{itemize}
    \item Iterate through the sorted list and compare each pair of adjacent elements.
    \item For each pair \texttt{(numbers[i], numbers[i+1])}, calculate the absolute difference.
    \item If the difference is less than the threshold, return \texttt{True}.
\end{itemize}

\textbf{4. Return Result:}
\begin{itemize}
    \item If no such pair is found after checking all adjacent pairs, return \texttt{False}.
\end{itemize}

\textbf{5. Complexity Consideration:}
\begin{itemize}
    \item Sorting the list takes $O(n \log n)$ time, and iterating through the list takes $O(n)$ time. Thus, the overall time complexity is $O(n \log n)$, which is efficient for large inputs.
\end{itemize}

\textbf{6. Precision Handling:}
\begin{itemize}
    \item Ensure that floating-point arithmetic is handled correctly by using Python's built-in functions which are designed to manage floating-point precision.
\end{itemize}

\vspace{0.5cm} \hrule \vspace{0.5cm}

\noindent\textbf{\Large Plan Simulation}

\textbf{Prompt:}
\begin{verbatim}
You are a programmer tasked with verifying a plan to solve a given problem 
using the **Python3** programming language.

[Problem and Plan shown above]

**Expected Output:**
Your response must be structured as follows:

### Simulation
- Take a sample input and apply plan step by step to get the output.
- Compare the generated output with the sample output to verify if your plan 
  works as expected.

### Plan Evaluation
- If the simulation is successful write **No Need to Modify Plan**.
- Otherwise write **Plan Modification Needed**.
\end{verbatim}

\textbf{LLM Response:}

\textbf{\#\#\# Simulation}

Let's simulate the plan with the provided test cases to verify its correctness.

\textbf{Test Case 1:} \texttt{numbers = [1.0, 2.0, 3.0]}, \texttt{threshold = 0.5}

\begin{enumerate}
    \item \textbf{Initial Checks:} The list has more than one element, so we proceed.
    \item \textbf{Sorting:} The list is already sorted: \texttt{[1.0, 2.0, 3.0]}.
    \item \textbf{Iterate and Compare:}
    \begin{itemize}
        \item Compare \texttt{1.0} and \texttt{2.0}: $|1.0 - 2.0| = 1.0$, which is not less than \texttt{0.5}.
        \item Compare \texttt{2.0} and \texttt{3.0}: $|2.0 - 3.0| = 1.0$, which is not less than \texttt{0.5}.
    \end{itemize}
    \item \textbf{Return Result:} No pairs found with a difference less than \texttt{0.5}, so return \texttt{False}.
\end{enumerate}

Output: \texttt{False} (Matches expected output)

\textbf{Test Case 2:} \texttt{numbers = [1.0, 2.8, 3.0, 4.0, 5.0, 2.0]}, \texttt{threshold = 0.3}

\begin{enumerate}
    \item \textbf{Initial Checks:} The list has more than one element, so we proceed.
    \item \textbf{Sorting:} Sort the list: \texttt{[1.0, 2.0, 2.8, 3.0, 4.0, 5.0]}.
    \item \textbf{Iterate and Compare:}
    \begin{itemize}
        \item Compare \texttt{1.0} and \texttt{2.0}: $|1.0 - 2.0| = 1.0$, which is not less than \texttt{0.3}.
        \item Compare \texttt{2.0} and \texttt{2.8}: $|2.0 - 2.8| = 0.8$, which is not less than \texttt{0.3}.
        \item Compare \texttt{2.8} and \texttt{3.0}: $|2.8 - 3.0| = 0.2$, which is less than \texttt{0.3}.
    \end{itemize}
    \item \textbf{Return Result:} A pair (\texttt{2.8}, \texttt{3.0}) is found with a difference less than \texttt{0.3}, so return \texttt{True}.
\end{enumerate}

Output: \texttt{True} (Matches expected output)

\textbf{\#\#\# Plan Evaluation}

The simulation of the plan with the provided test cases shows that the implementation works as expected. The plan correctly identifies whether there are any two numbers in the list that are closer to each other than the given threshold.

\textbf{No Need to Modify Plan}

\vspace{0.5cm} \hrule \vspace{0.5cm}

\noindent\textbf{\Large Code Generation}

\textbf{Prompt:}
\begin{verbatim}
You are a programmer tasked with solving a given problem using the **Python3** 
programming language. See the plan to solve the plan and implement code.

[Problem and Plan shown above]

--------
**Important Instructions:**
- Do not add any explanation.
- The generated **Python3** code must be inside a triple backtick (```) block.
\end{verbatim}

\textbf{LLM Response:}

\begin{verbatim}
from typing import List

def has_close_elements(numbers: List[float], threshold: float) -> bool:
    # Step 1: Initial check for minimal input
    if len(numbers) < 2:
        return False
    
    # Step 2: Sort the list
    numbers.sort()
    
    # Step 3: Iterate and compare adjacent elements
    for i in range(len(numbers) - 1):
        if abs(numbers[i] - numbers[i + 1]) < threshold:
            return True
    
    # Step 4: Return result if no close elements found
    return False
\end{verbatim}

\vspace{0.5cm} \hrule \vspace{0.5cm}

\noindent\textbf{\Large [I] Intermediate Simulation}

\textbf{Prompt:}
\begin{verbatim}
## Problem
[problem code]

## Code
[generated code]

## Your Task: Mental Execution
1. Select a sample input.
2. Trace the code execution step-by-step.
3. Track variable values.
4. Predict the final output.

If you find a logic error or mismatch with the plan, 
output: **CODE_SIMULATION_FAILED**
If the code seems correct, 
output: **CODE_SIMULATION_PASSED**
\end{verbatim}

\textbf{LLM Response:}

Let's select a sample input and trace the code execution step-by-step.

\textbf{Sample Input:}
\begin{verbatim}
numbers = [1.0, 2.8, 3.0, 4.0, 5.0, 2.0]
threshold = 0.3
\end{verbatim}

\textbf{Code Execution:}

\textbf{1. Initial Check:}
\begin{itemize}
    \item The code checks if the length of \texttt{numbers} is less than 2.
    \item \texttt{len(numbers) = 6}, so the check fails, and we proceed to the next step.
\end{itemize}

\textbf{2. Sorting the List:}
\begin{itemize}
    \item The list \texttt{numbers} is sorted.
    \item Sorted \texttt{numbers = [1.0, 2.0, 2.8, 3.0, 4.0, 5.0]}.
\end{itemize}

\textbf{3. Iterate and Compare Adjacent Elements:}

We iterate over the sorted list and compare adjacent elements.

\begin{itemize}
    \item \textbf{Iteration 1} (\texttt{i = 0}):
    \begin{itemize}
        \item Compare \texttt{numbers[0]} and \texttt{numbers[1]}: $|1.0 - 2.0| = 1.0$
        \item $1.0 \geq 0.3$, so continue to the next iteration.
    \end{itemize}
    
    \item \textbf{Iteration 2} (\texttt{i = 1}):
    \begin{itemize}
        \item Compare \texttt{numbers[1]} and \texttt{numbers[2]}: $|2.0 - 2.8| = 0.8$
        \item $0.8 \geq 0.3$, so continue to the next iteration.
    \end{itemize}
    
    \item \textbf{Iteration 3} (\texttt{i = 2}):
    \begin{itemize}
        \item Compare \texttt{numbers[2]} and \texttt{numbers[3]}: $|2.8 - 3.0| = 0.2$
        \item $0.2 < 0.3$, so the condition is satisfied, and the function returns \texttt{True}.
    \end{itemize}
\end{itemize}

\textbf{4. Return Result:}
\begin{itemize}
    \item Since we found two numbers (\texttt{2.8} and \texttt{3.0}) that are closer than the threshold (\texttt{0.3}), the function returns \texttt{True}.
\end{itemize}

\textbf{Conclusion:} The code correctly identifies that there are two numbers in the list that are closer to each other than the given threshold. The logic and execution match the expected behavior.

\textbf{Output: CODE\_SIMULATION\_PASSED}

\vspace{0.5cm} \hrule \vspace{0.5cm}

\noindent\textbf{\Large [O] Oracle-based Assumption Breaking (Round 1/3)}

\textbf{Prompt:}
\begin{verbatim}
You are an expert software tester (Red Team).
Your goal is to break the following code by finding hidden assumptions.

## Problem
[problem code]

## Code
[generated code]

## Your Task
1. Identify a weak assumption (Type, Value, Structure, or Relationship).
2. Produce a **Python test script** that calls the target function with a 
   breaking input.
3. Include an assert that would fail if the assumption is violated.

Format:
Assumption: <short text>
Test Script:
```python
result = <call>
assert <oracle about result>
```
\end{verbatim}

\textbf{LLM Response:}

\textbf{Assumption:} The function assumes that the input list \texttt{numbers} contains only valid floating-point numbers and does not handle special floating-point values like \texttt{NaN} (Not a Number) or \texttt{inf} (infinity), which can lead to unexpected behavior.

\textbf{Test Script:}
\begin{verbatim}
# call the function defined above
result = has_close_elements([1.0, float('nan'), 3.0], 0.5)
assert result == False, "The function should return False when the list 
contains NaN, as NaN comparisons are not valid."

# call the function defined above
result = has_close_elements([1.0, float('inf'), 3.0], 0.5)
assert result == False, "The function should return False when the list 
contains infinity, as infinity cannot be close to any finite number."
\end{verbatim}

In this test script, we are testing the function with lists that include \texttt{NaN} and \texttt{inf}. The function should ideally handle these cases gracefully, but due to the nature of these special floating-point values, the current implementation may not behave as expected. The assertions are designed to fail if the function does not correctly handle these cases.

\vspace{0.5cm} \hrule \vspace{0.5cm}

\noindent\textbf{\Large [L] Live Execution + Judge}

\textbf{Judge Prompt:}
\begin{verbatim}
You are an impartial Judge.

## Problem
[problem code]

## Proposed Test Case
[test script from Red Team]

## Task
Determine if this test case is VALID and CORRECT for the given problem.
1. Does the input satisfy all constraints (e.g. range, type, format)?
2. Is the asserted output (if any) logically correct?
3. Is it a fair test?

If the test case is valid, output: **VALID**
If it violates constraints or expects wrong output, output: **INVALID**
\end{verbatim}

\textbf{Judge Response:}

To determine if the proposed test cases are valid and correct for the given problem, let's analyze each aspect:

\textbf{1. Input Constraints:} The function \texttt{has\_close\_elements} accepts a list of floats and a float threshold. The proposed test cases use lists containing \texttt{float('nan')} and \texttt{float('inf')}, which are valid float values in Python. Therefore, the input satisfies the type constraints.

\textbf{2. Logical Correctness of the Asserted Output:}
\begin{itemize}
    \item For the test case with \texttt{float('nan')}: In Python, any comparison with \texttt{NaN} (Not a Number) is always false, including equality and inequality checks. Therefore, the presence of \texttt{NaN} should not affect the determination of whether any two numbers are closer than the threshold. The function should return \texttt{False} as no valid comparisons can be made with \texttt{NaN}.
    \item For the test case with \texttt{float('inf')}: Infinity is not a finite number, and it cannot be close to any finite number within a finite threshold. Therefore, the function should return \texttt{False} as no finite number can be within a finite threshold of infinity.
\end{itemize}

\textbf{3. Fairness of the Test:} The test cases are fair as they check the function's behavior with special float values (\texttt{NaN} and \texttt{Infinity}), which are edge cases that can occur in floating-point computations.

Given these points, the proposed test cases are logically correct and do not violate any constraints. They are valid tests for the function's behavior with special float values.

\textbf{Verdict: VALID}

\vspace{0.3cm}

\textbf{Live Execution Result:}
\begin{verbatim}
Executing test script with generated code...

def _concrete_verify_script(code, test_script, timeout=5):
    full_code = f"""
    {code}
    {test_script}
    """
    try:
        exec(full_code, exec_globals)
        return "PASS"
    except AssertionError:
        return "FAIL_ASSERT"
    except Exception:
        return "FAIL_CRASH"

Result: PASS (no assertion errors, no crashes)
\end{verbatim}

\vspace{0.5cm} \hrule \vspace{0.5cm}

\noindent\textbf{\Large Final Result}

\begin{center}
\fbox{\Large\textbf{PASSED after self-check}}
\end{center}

\vspace{3mm}

\begin{tabular}{ll}
\textbf{Total API Calls:} & 6 \\
\textbf{Total Time:} & 20.8 seconds \\
\textbf{Result:} & Solved (1/1 = 100\%) \\
\end{tabular}

\vspace{5mm}

The generated code correctly handles all edge cases including \texttt{NaN} and infinity without requiring any fixes. This demonstrates how SolidCoder's Live Execution component grounds verification in runtime reality rather than relying on mental simulation alone. The Judge correctly validated the Red Team's test cases as fair, and the Live Execution confirmed the code's robustness.

\end{tcolorbox}

\end{document}